\documentstyle[prl,aps,epsf, amsfonts]{revtex}

\def\8{\infty}

\def\undertext#1{\vtop{\hbox{#1}\kern 1pt \hrule}}

\def\VEV#1{\left\langle\,#1\,\right\rangle}

\def\dbyd#1#2{\frac{d#1}{d#2}}

\def\pbyp#1#2{\frac{\partial#1}{\partial#2}}

\def\br{\\ \nonumber & &}

\def\be{\begin{equation}}
\def\ee{\end{equation}}
\def\bea{\begin{eqnarray} & &}
\def\eea{\end{eqnarray}}
\def\ct#1{\cite{#1}}
\def\rf#1{(\ref{#1})}

\title{
Quantum Hall Transition in the Classical Limit}

\author { V. Gurarie$^{a,b}$ and A. Zee$^{a}$}

\address{$^a$Institute for Theoretical Physics, University of California,
Santa Barbara CA 93106-4030}
\address{$^b$Department of Theoretical Physics, Oxford University, 1 Keble
Road, Oxford OX1 3NP, United Kingdom}

\date{\today}

\begin {document}
\draft
\maketitle

\begin{abstract}

We study the quantum Hall transition using the
density-density correlation function. We show that in the limit
$\hbar \rightarrow 0$ the electron density moves along the percolating
trajectories, undergoing normal diffusion. The localization exponent
coincides with its percolation value $\nu={4\over 3}$. The framework
provides a natural way to study the renormalization group flow from
percolation to quantum Hall transition. We also confirm numerically
that the critical conductivity of a classical limit of quantum 
Hall transition is $\sigma_{xx} = {\sqrt{3} /4}$.

\end{abstract}

\section{Introduction}

Integer quantum Hall transition is a localization-delocalization
transition experienced by a quantum mechanical particle moving in
a two dimensional plane in a magnetic field perpendicular to the
plane and in a random scalar potential $V(x)$. As the energy of the
particle approaches its critical value, 
the particle becomes more and more delocalized. 
In spite of the fact that the problem has been
studied numerically and experimentally for 20 years, it
has resisted exact theoretical descriptions.

In this paper we examine a new method to probe the
properties of the quantum Hall transitions recently
introduced by Girvin {\sl et al}
\ct{Girvin}. We show that this method provides a natural
framework in which the classical limit of quantum Hall transition 
(classical percolation)
can be studied. We show, both numerically and analytically, 
that even in this classical limit the percolating particle exhibits
normal diffusion, just as the quantum critical particle would. 
The critical exponent $\nu$ turns out to be equal to the percolation
exponent ${4 \over 3}$. In this framework, the renormalization group
flow from percolation to the quantum Hall transition can be studied. 

The method introduced in \ct{Girvin} is 
based on the introduction of the density-density correlation
function \be G(x,t) = \sum_n \VEV{n| \rho(x,t) \rho(0) |n} \ee
where $\rho$ is the density operator and the sum goes over the
Hilbert space of states.
This correlation function has a simple meaning. It is the
probability that a quantum particle, localized at the origin at
the initial moment in time, is found at the position $x$ at a time
$t$. Indeed, set up a wave function of a particle localized at the
origin, \be \psi(x) = \sum_n \psi_n^*(0) \psi_n(x), \ee where
$\psi_n$ are the eigenstates of the Hamiltonian of the problem. At
time $t$ the wave function will be (we set $\hbar=1$) \be
\psi(x,t) = \sum_n \psi^*_n(0) \psi_n(x) e^{i E_n t}, \ee where
$E_n$ are the eigenenergies.  The probability 
is equal to \be \label{prob} {\left|
\psi(x,t) \right|}^2 = \sum_{m,n} \psi^*_n(0) \psi_n(x)
\psi^*_m(x) \psi_m(0) e^{i \left( E_n- E_m \right) t}. \ee

The matrix elements of the density operator are \be
\rho_{m,n}(x)=\psi^*_m (x) \psi_n(x). \ee It is therefore obvious
that the correlation function \be \label{rr} G(x,t)= \sum_{m,n}
\rho_{m,n} (x,t) \rho_{n,m} (0) \ee is equal to the probability
\rf{prob}.

It sounds reasonable that such a correlation function, which tests how far
a particle can go in a given amount of time, is useful when
studying localization. In the localization literature, however, a
different function is usually studied. It is called the spectral
function and is given by the square of the single particle Green's
function \ct{Chalker}
\be \label{defS}
S(x, \omega; E)= 
%G\left( x,E+{\omega \over 2} \right) G^* \left(
%x,E-{\omega \over 2} \right) =
\sum_{m,n} {\psi^*_m (x) \psi_m(0)
\psi_n(x) \psi^*_n (0) \over \left( E-{\omega \over 2} - i
\epsilon-E_n \right) \left( E+{\omega \over 2}+ i \delta-E_m
\right) }, \ee with $\epsilon$ and $\delta$ infinitesimally small
numbers. As is well known, the conductance of the system is simply
related to the spectral function $S(x, \omega; E)$.  This function
contains more information than the correlation $G$. It tells us how far
a particle can go if it propagates along the state with a given
energy $E$. The correlation function $G$ is related to the integral of $S$
over $E$, \be \label{inte}
2 \pi G(x,\omega+i \epsilon+ i \delta) =
\int_{-\infty}^\infty dE \ S(x,\omega; E), \ee
where $G(x,\omega)$ is a Fourier transform of $G(x,t)$. The last formula can
be obtained by directly integrating \rf{prob} and \rf{defS}.

One could argue that the density-density correlator $G(x,t)$ we
just introduced would not give much information about
the localization properties. Indeed, in a transition such as Anderson
transition in 3D, this function will always be long ranged due to the 
contributions of the delocalized states to \rf{inte}.
In contrast, the spectral function $S(x,\omega; E)$
tests only the states at the energy $E$.

The insight of Girvin {\sl at al} was that in quantum Hall transition
only the centers of the Landau bands are believed to be
delocalized. Moreover, the physics of quantum Hall transitions
allows us to restrict the particle to the lowest Landau level. In
practical terms, that means restricting the sum in \rf{rr} to go
only over the states in the lowest Landau level. Then there is
only one delocalized state. In \rf{inte}
only the states whose localization length is larger
than $x$ contribute to the integral over all the energies. 
Consequently the density-density correlation function includes in
itself some useful information about quantum Hall transitions.

It follows from the scaling theory of phase transitions that the
localization length $\xi(E)$ is divergent as we approach the
localization transition with the exponent $\nu$, \be \label{loc}
\xi(E) \propto |E-E_c|^{-\nu} \ee where  $E_c$ is the critical
energy. It has also been argued \ct{Chalker} that a quantum
particle in 2D at the localization-delocalization transition
undergoes normal diffusion, which in particular results in a finite
critical conductivity. Heuristically people sometimes 
estimate the
Fourier-transformed 
density-density correlator by an interpolating method based on \rf{inte}.
%\be \label{int0} G(k,t) = \int
%dE \ f(k,t; E). \ee 
Until the particle hits the localization length, 
we assume that
\be
\label{int1} S(k,t; E) \propto \exp \left( - {D t { k^2 \over 4} } \right), \
{\rm if} \ Dt \ll \xi^2(E), \ee which describes diffusion with the
diffusion constant $D$. This assumption is true only if $k \rightarrow 0$
\ct{Chalker}. The diffusion at small $k$ reflects the existence of critical
conductivity in quantum Hall effect. 
On the other hand, at larger $k$ the motion of the particle becomes more
complicated.
That reflects the fact that critical conductance is not Ohmic and in fact
dependent on the shape of the sample. Obviously, \rf{int1} will not work
if the particle is deeply in the localized regime, when $|E-E_c| \gg E_{\rm max}
$, with $E_{\rm max}$ describing the scale where \rf{int1} and \rf{loc}
break down.

If we wait long enough, 
\be \label{int2} S(k,t; E) \propto  \exp \left( - { {k^2 \over 4}
\xi^2(E)} \right), \ {\rm if} \ Dt \gg \xi^2(E). \ee Roughly
speaking, $G(x,t)$ can be estimated as \be \label{int3}
G(k,t) \propto \int_{E_c-E_{\rm max}}^{E_c+E_{\rm max}} dE \ \exp
\left( - { k^2  \over 4 \left( {1 \over  Dt} + {1 \over \xi^2(E)} \right)} 
\right),\ee
where the function being integrated interpolates between the
limiting behavior \rf{int1} and \rf{int2}. Using 
\rf {loc} and changing the variables $E = E_c+
z t^{-{1 \over 2 \nu}}$ , we can estimate this integral to go as 
at \bea \label{int4} 
G(k,t) = t^{-{1\over 2
\nu}} \int_0^{t^{2 \nu} E_{\rm max}} 
dz \ \exp \left( - {  k^2 t \over 4 \left( D^{-1} + z^{2 \nu}
\right)} \right) = \br E_{\rm max}- t^{-{1\over 2
\nu}} \int_0^{t^{2 \nu} E_{\rm max}} 
dz \ \left[ 1- \exp \left( - {  k^2 t \over 4 \left( D^{-1} + z^{2 \nu}
\right)} \right) \right]
\propto  1-t^{-{1\over 2 \nu}} g(k^2 t), \eea 
where $g$ is some
complicated function. We used that the integral on the second line can
be extended to infinity. 
$G(x,t)$ is diffusive at small $k$ (a function of $x^2 \over t$)
with a prefactor which is $t$ dependent. Knowing the prefactor
allows us to determine the critical exponent $\nu$. This formula
has been tested numerically in \ct{Girvin} and was found to give
values of $\nu$ for the quantum Hall transition
around the value accepted by the community.

Imprecise as the derivation of this formula may seem, we will show
later in this paper how it can be given a precise meaning in the
classical limit. 

\section{Equation of Motion}

Now we would like to demonstrate that, as was first shown in
\ct{Girvin}, the density-density correlation function $G(x,t)$,
when projected to the lowest Landau level and in the presence of a
scalar potential $V(x)$, satisfies a simple equation.

Choose the basis of states in the lowest Landau level as in  
(see \ct{Itsykson} and references therein) \be
\label{efunc}  \psi_n (z,z^*) = {z^n \over \sqrt{ {\left( 2 l^2
\right)}^n n! \pi }} \exp \left( - {z z^* \over 4 l^2} \right),
\ee where, as always, $z=x+i y$, complex coordinates on the 2D
plane and $l$ is the so called magnetic length.  
Consider the Fourier transformed density operator \be
\rho_k=\exp \left[ {i \over 2} \left( k z^*+ k^* z \right)
\right], \ee where similarly to the complex coordinates
$k=k_x+i k_y$.
Let us project this operator to the lowest Landau
level, following the methods developed in \ct{Girvin1}. The
operator consists of the product of two factors. One of them,
$\exp \left[ {i \over 2} k^* z \right]$ is analytic in $z$ and so
acts entirely within the lowest Landau level. As such, it does not
have to be projected. Indeed, if we multiply one of
the wave functions \rf{efunc} of the lowest Landau level by this factor, the
result can be reexpressed as a linear combination of other
functions of \rf{efunc}. The other factor involves $z^*$ and
therefore does not act within the lowest Landau level. It has
to be projected. We note, however, that the operator \be
\label{zbar} \hat z = 2 l^2 {\partial \over
\partial z} + {1 \over 2} z^* \ee acts within the lowest Landau
level, \be \hat z \psi_n(z,z^*) = \sqrt{ 2 l^2
n  } \ \psi_{n-1} 
  \ee 
Furthermore, 
\be
\int d^2z \ {\left(z^*\right)}^n  \left(z^*\right)^s z^m \exp \left(-{z z^* \over
2 l^2 } \right) = \int d^2 z \ {\left(z^*\right)}^n \exp
\left( - {z z^* \over 4 l^2} \right) \left( 2 l^2 
{\partial \over \partial z} + {1 \over 2}
z^* \right)^s
\left[ z^m \exp \left(-{z z^* \over 4 l^2 } \right) \right]. \ee
In other words,
\be
\VEV{n|\left(z^*\right)^s|m}=
\VEV{n\left| \left( 2 l^2 {\partial \over
\partial z} + {1 \over 2} z \right)^s \right|m }
\ee

 This
allows us to write down the projected density operator as in
\be
\label{de} \rho_k  =\exp \left[ {i \over 2} k \hat z \right]
\exp \left[ { i \over 2}
k^*
z  \right]. \ee

Normally the density operators commute with each other, but the
projected density operators do not. Multiplying
two density operators together and using \rf{zbar} we obtain
\be \rho_k \rho_q = \exp \left[ {l^2 \over 2} \left(  k \cdot  q +
i k \times q \right) \right] \rho_{k+q}, \ee where $k \cdot q$ is
the dot and $k \times q$ is the cross products of two vectors $(k_x,k_y)$
and $(q_x,q_y)$.

Finally we note that the Hamiltonian of a particle moving in a
lowest Landau level in the presence of the scalar potential $V(x)$
is just that potential itself projected to the lowest Landau
level, \be H= \int dx \ V(x) \rho(x) = \int dk \ V(k) \rho_{-k} .
\ee Therefore, the equation of motion for the operator $\rho$ is
\be \label {master} i \hbar {\partial \over
\partial t} \rho_k =  \left[ H, \rho_k \right]= \int d^2q \ 2 i \sin
\left( {l^2 \over 2} k \times q \right) V(k-q)\exp \left[ - {l^2
\over 2} \left(k^2- k \cdot 
q\right) \right] \rho_q, \ee where we restored the
$\hbar$ factor. Clearly $G(k,t)={\rm tr} (
\rho_k(t) \rho_{-k}(0))$ satisfies 
\be \label {master1} i \hbar {\partial \over
\partial t} G(k,t) = \int d^2q \ 2 i \sin \left( {l^2 \over 2} k
\times q \right) V(k-q)\exp \left[ - {l^2 \over 2}
\left(k^2-k \cdot q\right) \right] G(q,t), \ee 

with the boundary condition
\be G(k,0) = {\rm tr} \left( \rho(k) \rho(-k) \right) = N \exp 
\left( - {l^2 \over 2} k^2 \right),
\ee N being the number of states in the lowest Landau level. 
Solving this equation allows
us in principle to determine $G(k,t)$. Therefore, it encodes all the
properties of quantum Hall transitions. Unfortunately, while this
equation is simple to write down, solving it is not a trivial
task.

\section {Classical Limit}

Let us consider the classical limit of the quantum Hall
transition.  The classical limit $\hbar \rightarrow 0$ is
equivalent here to the limit $l \rightarrow 0$ since 
$l^2 \propto \hbar$.
In this limit, the equation \rf{master1} becomes \be \label{macl}
\pbyp {G(x,t)}{t} = {l^2 \over \hbar} \epsilon_{i j}
\partial_i V
\partial_j G(x,t). \ee
Since $l^2 \propto \hbar$, the planck constant completely drops out.
In what follows, we choose the units where $\hbar / l^2=1$. 

The formal solution to this equation is straightforward to obtain:
\be
\label{Gcla}  G(x,t) = \delta(x(t)-x), \ee where $x(t)$ is a
trajectory satisfying the following equation \be \label{tra}
\dbyd{x_i(t)}{t}= \epsilon_{ij} \partial_j V(x(t)). \ee Studying
the solutions to the equation \rf{macl} reduces to studying the
trajectories satisfying \rf{tra}.

\begin{figure}[tpb]
\centerline{\epsfxsize=2in \epsfbox{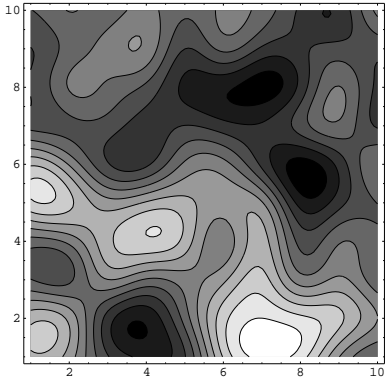}}
\caption{Equipotential lines of a typical random potential $V(x)$. These
are trajectory lines of the solutions to \rf{tra}. The potential in this 
example
is generated as a sum of Gaussian shapes located at the nodes of a square 
lattice
with random
amplitudes. }
\label{Fig4}
\end{figure}

The tangent to the trajectory $x(t)$ is always perpendicular to
the gradient of the potential $V(x(t))$. Therefore,
$V(x(t))$ is conserved along the trajectory. In other words,
the particle moves along the equipotential lines, as shown on
Fig. \ref{Fig4}. The value of the
potential along the trajectory becomes an important parameter
$V_0=V(x(t))$ which can be assigned to each trajectory satisfying
\rf{tra}. In a way, the potential $V(x)$ can be thought of as a 
mountainous
landscape with peaks and valleys. As water fills the valleys of the landscape
to the height $V_0$, the equipotential lines of Fig \ref{Fig4} are the shapes 
of the
lakes formed in this way.

In this form the problem we study becomes equivalent to
classical percolation \ct{percolation}. 
Indeed, the potential $V(x)$ is random and independent in
different points in space. We can consider the value of the
potential at a point $x$ as a random variable. Take a
particular trajectory $x(t)$ with a particular value of
$V(x(t))=V_0$. Let us say that the point $x$ is ``on" if
$V(x)>V_0$ and ``off" otherwise. Then the trajectory surrounds the
clusters of the ``on" sites. The trajectory becomes the so-called
``hull" of the percolation cluster, and the area it surrounds is
the percolation cluster itself (or the islands in the landscape analogy,
for $V_0>0$). In the percolation problem, the
site is ``on" with a probability $p$. In our problem it can be
expressed as in \be p=\int_{V_0}^\infty P(V) \ dV \ee where $P(V)$
is the probability for the values of the potential at different
points in space (sites). In this way, we recover the classical
percolation picture of the quantum Hall transitions \ct{Trudgman}.
The electron density percolates along the equipotential lines of
the scalar potential. The trajectories are closed. However, as 
$V_0 \propto \left( p-{1\over 2} \right)$ approaches zero, 
the trajectories become
more and more extended in space. Denoting by $\xi$ the root mean square
distance between points on the trajectory, we say that the percolation
hulls exhibit the delocalization transition,
with $\xi \propto V_0^{-\nu}$. 

Define the area of a cluster as
\be
A=\int d^2 x \ \theta(V(x)-V_0), \ee
$\theta(x)=1$ when
$x>0$ and $\theta(x)=0$ when $x<0$.

It turns out that the time it takes to go along the boundary of a cluster 
is just $dA/dV_0$, or
\be \label{time}
T=\int d^2 x \ \delta(V_0-V(x)). \ee This can be derived directly
from \rf{tra}. Alternatively, one can argue that \rf{tra} are
the Hamiltonian equations, with $V$ being the Hamiltonian. Then
$S$ is the ``adiabatic invariant'' and its derivative is obviously time. 

From here one can conclude that, since the area of a cluster goes as
$A \propto V_0^{-\gamma}$, the time goes as $T \propto V_0^{-\gamma-1}$,
where
$\gamma={43 \over 18}$ is another well known percolation exponent 
\ct{percolation}.
However, as we will see below, $T$ is not equal to the time it takes
for a particle to go around the cluster. Rather it is much bigger
having to do with the ``swiss cheese'' shape of a typical cluster.
As a result, $T$ sums the time it takes go around the cluster, and
around each hole in the cluster.

The classical percolation  picture of the quantum Hall transitions
was suggested a long time ago. It was subsequently demonstrated
that it does not capture the physics of quantum Hall transitions.
Indeed, the exponent $\nu$ computed in this way coincides with the
its value for percolation $\nu={4 \over 3}$ \ct{Nijs}, while the
true value for $\nu$ is believed to be around $\nu=2.35\approx {7 \over 3}$. 
We can
say that the classical density-density correlation function $G(x)$
forgets about the phase a quantum particle carries with itself. It
was also known that the classical percolation picture can be
derived as a classical limit of the Chalker-Coddington model \ct{CC},
which is believed to lie in the same universality class as quantum Hall
transitions. 

Some time ago, F. Evers undertook a 
thorough numerical study of the solutions to the equations \rf{tra}
at zero energy $V_0$, motived by classical limit of quantum Hall
transition,  
in the paper \ct{Evers}. Its results will be useful for us
below.  

Now we are going to study percolation
as a classical limit of quantum Hall transition. 
The approach we are
using allows us to give a precise definition of what
percolation is for quantum Hall transition, which quantum
Hall quantities percolate in the classical picture and finally how
they change as we allow for finite $l$ to account for the quantum
effect.

\section {Percolation}

To study the percolation limit, we suggest a lattice model of the
equation \rf{tra}. As a possible lattice model, we could have
taken the classical limit Chalker-Coddington model \ct{CC}. This is the
version of the model where a particle turns left or right at each
node with a certain probability, instead of a certain quantum probability
amplitude as in the original work \ct{CC}. However, to make a closer
connection with percolation, we would like to consider a somewhat different
model.

\begin{figure}[tpb]
\centerline{\epsfxsize=1.5in \epsfbox{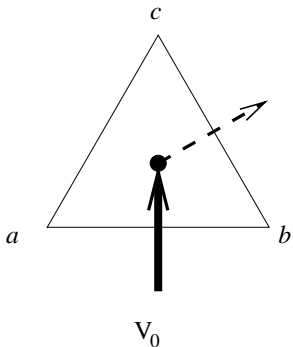}}
\caption{The particle moves in such a way that the potential $c$ of the left node
of the bond the particle crosses
is greater than $V_0$ which in turn is greater than the potential $b$
of the right node.}
\label{Fig1}
\end{figure}

\begin{figure}[tpb]
\centerline{\epsfxsize=4.0in \epsfbox{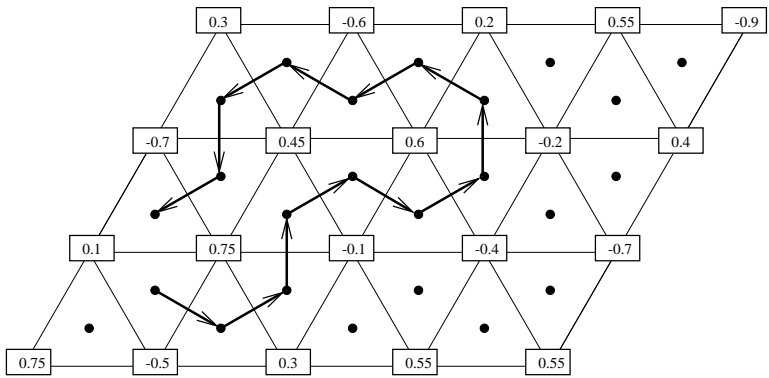}}
\caption{A typical trajectory. The value of its potential $V_0=0.4$}
\label{Fig2}
\end{figure}

Consider a triangular lattice. Associate with
each of its sites a random number $V$ drawn from a probability
distribution $P(V)$. Now associate the points of the dual
lattice with the path $x(t)$. Each path is assigned a number $V_0$, the value
of the energy on this particular path. At each step the path
moves in such a way  that 
the potential  of the left node
of the bond the particle crosses
is greater than $V_0$ which in turn is greater than the potential
of the right node. For example, the particle on Fig \ref{Fig1} enters the
triangle through the horizontal bond, as shown by the solid line. 
Therefore, $a > V_0 > b$. If $c>V_0$, it would leave
the triangle through the right bond, as shown by the dashed line.
If $c<V_0$, then it would leave the triangle through the left bond.
The trajectory then becomes the hull of percolation clusters. A typical 
hull is shown on Fig \ref{Fig2}.

The equation \rf{macl} also contains dynamical information. 
The velocity of the particle which follows \rf{tra} is proportional
to the gradient of the potential. We model it by saying that the time it
takes for a particle to traverse a bond of the triangular lattice 
is inversely proportional to the potential difference at the ends
of the bond. For example, the time it takes to traverse the bond
of the Fig. \ref{Fig1} (the process denoted by the 
dashed line) is $1/(c-b)$. 

Using this model, we performed extensive numerical simulations.  
In the simulations we assumed that the values of the potential at the
nodes of the triangular lattice are random numbers uniformly distributed
on the interval $[-1,1]$. The lattice size was 8500 by 8500, and
we discarded the trajectories which went outside the lattice. 
The typical trajectories were, however, of much smaller size, about 500. 
We discovered
that the time it takes for a particle to go along a trajectory 
is essentially proportional to the trajectory's length. 
This is because the velocity of the trajectory remains roughly the
same as $V_0$ is varied, having no singularity or zero as $V_0$ passes
through 0. Indeed, it is easy to estimate the average velocity by assuming
that all the potentials at the corners of a triangle
are random independent numbers. Then all we need to do is to average
$1/(V_1-V_2)$ where $V_1$ and $V_2$ are the values of the potential
at the ends of the bond the trajectory is traversing. For the potential
uniformly distributed on the interval $[-1,1]$ we obtain that the
average velocity is very weakly $V_0$ dependent, with its value
being $\log(4)$ when $V_0=0$. This was confirmed numerically.
In fact, that coincides with the earlier observation in \ct{Evers}. 

Interestingly, that allows us to find the time $T$ it takes for a particle
to go around the cluster. The length of a trajectory is known to
go as 
\be L \propto V_0 ^{- \gamma'}, \ee
where $\gamma'=2$ \ct{Saleur}. Since the average velocity is only weakly 
$V_0$ 
dependent, we obtain that $T \propto V_0^{-2}$. That should be contrasted
with the time it takes to go around the cluster and around each hole
in the cluster which we found in the previous section.

\begin{figure}[tpb]
\centerline{\epsfxsize=5in \epsfbox{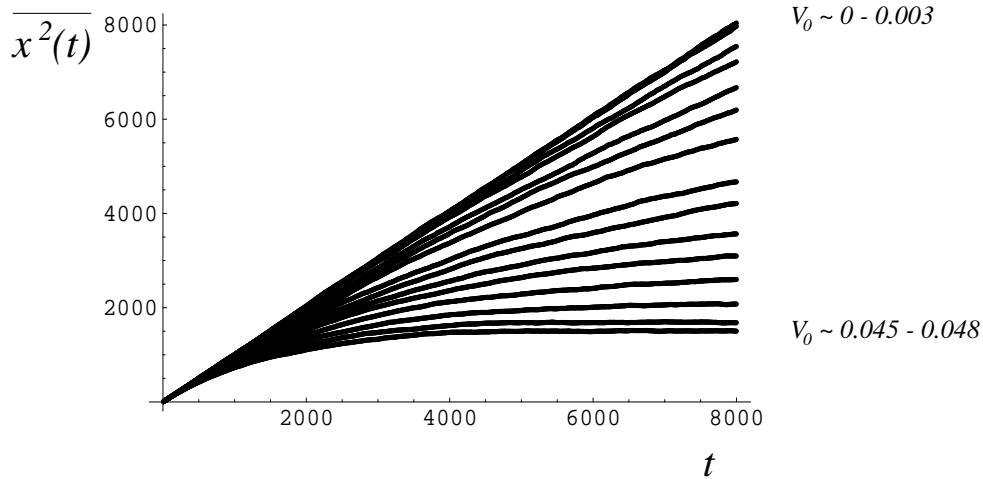}}
\caption{$\overline{x^2(t)}$ computed for trajectories with various
$V_0$. The highest line represents the lowest $V_0$. The potential
at the nodes of the lattice is uniformly distributed from $-1$ to $1$. Time
$t$ is simply the number of steps. $x^2$ is measured in units where the
triangular lattice size of Fig. \ref{Fig2} is one.}
\label{Fig3}
\end{figure}

Replacing time with length we studied $\overline {x^2(t)}$ where
the average means averaging over different trajectories with the
same value of $V_0$. In our simulations we averaged over about $10^6$
trajectories. Fig. \ref{Fig3} summarizes the results of the
numerics.

It is clear that $\overline{x^2(t)}$ is a straight line 
\be 
\label{diffu}
{\overline{x^2(t)}} = D  t, \
t \ll t^*
\ee
until a certain time $t^*$ which varies with $V_0$. After that,
\be
\label{local}
\overline{x^2(t)} = \xi^2, \ t \gg t^*,
\ee with the parameter $\xi$ depending on 
$V_0$ but not on time. 

The length 
$\xi$ is none other than the localization (correlation) length of the
percolation problem we introduced above. 
Indeed, if we wait long enough, the particle
satisfying \rf{tra} starts winding around the closed trajectories such as
on Fig. \ref{Fig4}. Then $\overline{x^2(t)}$ measures
the mean square
distance between two points on the percolation cluster's hull
which is the trajectory we study. So we can
write down 
\be
\xi \propto V_0^{-{4 \over 3}},
\ee
which we also confirmed by our numerical simulations. 

The numerical result that $\overline{x^2(t)} \propto t$ for
$t<t^*$ shows that the particle undergoes a normal diffusion. While
this is what is expected of a quantum particle in a quantum Hall transition
\ct{Chalker}, it is far less obvious that a classical particle 
should do the same. It is also not obvious how to explain this from
a purely percolation point of view. 

This result was previously obtained in \ct{Evers} where $\overline{x^2(t)}$
for the trajectories at $V_0=0$ was studied. 
There it was shown
that the diffusion was due to a ``mysterious'' cancellation of exponents in 
percolation. Perhaps, a certain scaling relation between those 
exists which leads to the cancellation, 
but at the moment we are not aware of it.

Now 
we would like to note that a convenient way to compute $\nu$ would be just
summing all the trajectories of Fig. \ref{Fig3} together. As above, we expect
this to be
\be
\sum_{V_0} \overline{x^2(t)}= 
\int dV_0 \ P(V_0) {1 \over {1 \over Dt} + V_0^{2 \nu}}
\propto t^{1 - {1 \over 2 \nu}}
\ee
Numerically this exponent is around $0.62$ which corresponds to $\nu=1.30$,
very close to the exact value. This method proves to be very convenient
when studying quantum Hall transition in the quantum regime as well \ct{BG}. 

Knowing $x(t)$ it is not difficult to compute $G(x,t)$ according to \rf{Gcla}.
\be G(k,t) = \sum_{V_0} \overline {\exp \left( i k x(t) \right) }, \ee
so that the average is taken over all the trajectories.
Compare this formula with \rf{inte}. We are led to conjecture that
\be S(x,t; V_0) = 2 \pi P(V_0) \ 
\overline {\exp \left( i k x(t) \right)}, \ee
where the average, as always,  
is taken over the trajectories with the same $V_0$. Indeed,
not only \rf{inte} is automatically satisfied, but also this form of 
$S$ reflects
the fact that the particle propagates with a given energy. 

With this definition of $S$ and $G$ it is easy to check that the
formulae \rf{int1}-\rf{int4} are indeed true and precisely defined.
Indeed, their applicability was at small $k$ where
\be
\overline{ \exp \left( i k x(t) \right)} \approx 1 - {k^2 \over 4}
\overline {x^2(t)}
\approx \exp \left( - {k^2 \over 4} \overline{x^2(t)} \right).\ee
From this point on, \rf{int1}-\rf{int4} can be carried on without
any further obstacles. The only difference between classical and quantum
case will be the value of $\nu$, which in the classical case is equal to
$4/3$. 

It is interesting to note that at large $k$ 
the functions $S$ and $G$ will be very different in quantum and classical
case. In the quantum case, $S$ is a function of the ratio $q^2/\omega$.
In the classical case this question was investigated in \ct{Evers}. There 
it was shown that $S(x,t; 0)$, computed at zero energy, 
was a function of $q^{7 \over 4} /\omega$,
with $7/4$ being the fractal dimension of the infinite percolating
cluster. We refer the reader to \ct{Evers} for further information on
this subject. 
In the meanwhile, the form of $G$ or even $S$ at generic values of
energy in the classical case remains unknown. 

Finally, we would like to comment on the interesting feature of the
numerical data of Fig. \ref{Fig3}. The diffusion constant, when read off
this data (given in units where the bonds of the triangular lattice 
of Fig. \ref{Fig2} have
unit length), is equal to 
\be D=1.008 \pm 0.014 \ee
This number is remarkably close to $1$

Assuming that $D=1$
is indeed an exact relation, we can demonstrate that the critical conductivity
is equal to
\be \label{sigma} \sigma_{xx} = {\sqrt{3} \over 4} {e^2 \over h},
\ee an exact result recently derived in \ct{Cardy} by a very different 
approach. 
To do that, we need to use the relationship between the diffusion constant
and the critical conductivity proposed by F. Evers and W. Brenig
in \ct{Evers1}. 

According to the Einstein's relation,
the conductivity is given by
\be \label{einst} \sigma_{xx} = e^2 \rho {D \over 4}, \ee
where $e$ is the electron charge and $\rho$ is the density of states per unit
area per unit energy at the transition. The problem is to find $\rho$ in the
lattice model introduced here. 

We can use a standard quasiclassical argument to find the density of states
for a particle moving in a magnetic field and a potential $V(x)$.
The total number of states whose energy is below $E$ is given, in the
quasiclassical limit, by the area of the region where $V(x)<E$, divided
by $2 \pi l^2$,
\be
A(E)= {1 \over 2 \pi l^2} \int d^2 x \ \theta \left( E-V(x) \right).
\ee
The density of states per unit area is
\be
\label{rhoE}
\rho(E) = {1 \over 2 \pi l^2 A} \int d^2 x \ \delta \left(E-V(x) \right),
\ee
where $A$ is the total area.

Comparing $\rho(E)$ with \rf{time} we see that $\rho(E)$ coincides with the
time $T$ it takes for a particle to go over all the trajectories with 
energy $E$. Restoring $\hbar$ and $l$ dependence in \rf{time} we find
\be
\rho(E) = {T \over h A} 
\ee
Therefore, the conductivity reduces to
\be
\label{ssigma}
\sigma = {e^2 \over h} {D T \over 4 A}
\ee
 It is now clear that the overall answer
is independent of the units in which we measure $D$, as long as they are
the same as the ones for $A$ and $T$.

In the lattice model of Fig. \ref{Fig2} the total time is equal to the
total number of links available for a particle with zero energy,
the energy of the transition. 
Assume that the total number of triangles
is equal to $N$. The area $A$ of the lattice is equal to
$N \times {\sqrt{3} \over 4}$, where ${\sqrt{3} \over 4}$ is the area
of a single triangle. The total number of links is $L={3 N \over 4}$. 
We took into account that each triangle
has 2 links with a probability $3/4$.
This is because with a probability
of $1/4$ all three energies at the vertices of a triangle are either less
or greater than 0, in which case there are no trajectories in the triangle
at all. But each link belongs to two triangles, hence the overcounting 
and the answer of ${3 N \over 4}$. Substituting this into \rf{ssigma}
and taking into account that $D \approx 1$, we arrive at

\be \sigma_{xx} = D {\sqrt{3} \over 4} {e^2 \over h} \approx {
\sqrt{3} \over 4} {e^2 \over h}, \ee in very good agreement with \ct{Cardy}.

This result is also in agreement with the one of \ct{Evers1} where the critical
conductivity was estimated at $\sigma_{xx} \approx 0.45 {e^2 \over h}$.

Notice that the conductance of a finite sample would not be Ohmic
since the particle does not undergo a simple random walk, as reflected
in a complicated structure of $G$ or $S$ at finite momenta $k$. Only
for very thin samples, when it takes little time for a particle to
reach one end of a sample from another, can the Ohm's law be applicable,
in agreement with \ct{Cardy}.

\section{Conclusions}

To summarize, we showed that in the classical limit the density of
a particle undergoing the quantum Hall transition moves along the
percolation hull trajectories. At small times this particle exhibits
normal diffusion, while at large times it localizes with the localization
length exponent given by the percolation $\nu={4 \over 3}$. Our paper
sets a new framework in which the classical percolation to quantum Hall
effect renormalization group flow can be studied. Finally, we confirm
numerically with a high degree of accuracy the result $\sigma_{xx} = \sqrt{3}/4$
for the 
critical conductivity of a classical quantum Hall transition. 

\section{Acknowledgments}

The authors are grateful to S. Boldyrev for many useful discussions and 
comments, to F. Evers who
painstakingly explained his method \ct{Evers1}
of relating the lattice diffusion constant
to conductivity to us, to J. Cardy for interesting comments and to 
to E. H. Rezayi for making his computer 
facilities available to us. 

This
work was supported by the NSF grant PHY 94-07194

\end{document}